\documentclass[10pt,journal,compsoc]{IEEEtran}
\ifCLASSOPTIONcompsoc
  \usepackage[nocompress]{cite}
\else
  \usepackage{cite}
\fi
\ifCLASSINFOpdf
\else
\fi
\usepackage{setspace}
\usepackage{graphicx}
\usepackage{wrapfig}
\usepackage{epstopdf}
\usepackage{subcaption}
\usepackage{caption}
\usepackage{multirow}
\usepackage{nicefrac}
\usepackage{booktabs}
\usepackage{amsmath}
\usepackage{amsthm}
\usepackage[algo2e]{algorithm2e} 
\usepackage{algorithm}
\usepackage{verbatim} 
\usepackage{amssymb}
\usepackage{bbm}
\usepackage[english]{babel}

\begin{document}
\title{Influence maximization on temporal \\networks: a review}

\author{Eric Yanchenko, 
        Tsuyoshi Murata 
        and~Petter Holme% <-this % stops a space
\IEEEcompsocitemizethanks{\IEEEcompsocthanksitem E. Yanchenko: Department of Statistics at North Carolina State University, USA, and Department of Computer Science at Tokyo Institute of Technology, Japan. E-mail: ekyanche@ncsu.edu \protect
\IEEEcompsocthanksitem T. Murata: Department of Computer Science at Tokyo Institute of Technology, Japan.

\IEEEcompsocthanksitem P. Holme: Department of Computer Science at Aalto University, Finland and Center for Computational Social Science at Kobe University, Japan.

}% <-this % stops an unwanted space
}

\markboth{Journal of \LaTeX\ Class Files,~Vol.~14, No.~8, August~2015}%
{Yanchenko \MakeLowercase{\textit{et al.}}: Influence maximization on temporal networks: a review}

\IEEEtitleabstractindextext{%
\begin{abstract}
Influence maximization (IM) is an important topic in network science where a small seed set is chosen to maximize the spread of influence on a network. Recently, this problem has attracted attention on temporal networks where the network structure changes with time. IM on such dynamically varying networks is the topic of this review. We first categorize methods into two main paradigms: single and multiple seeding. In single seeding, nodes activate at the beginning of the diffusion process, and most methods either efficiently estimate the influence spread and select nodes with a greedy algorithm, or use a node-ranking heuristic. Nodes activate at different time points in the multiple seeding problem, via either sequential seeding, maintenance seeding or node probing paradigms. Throughout this review, we give special attention to deploying these algorithms in practice while also discussing existing solutions for real-world applications. We conclude by sharing important future research directions and challenges.

\end{abstract}

}

\maketitle

\IEEEdisplaynontitleabstractindextext

% \fi
\IEEEpeerreviewmaketitle


\section{Introduction}

Networks, or graphs, are a simple tool to abstractly represent a system involving interacting entities, where the objects are modeled as nodes and their relationship as edges. Because of their generality and flexibility, many real-world settings have leveraged networks over the past few decades including: online social networks~\cite{garton1997studying, mislove2007measurement, phuvipadawat2010breaking}, infrastructure networks~\cite{latora2005vulnerability, liu2020review, guimera2004modeling} and biological process networks~\cite{girvan2002community, pavlopoulos2011using}. Recently, there has been great interest in not only understanding the topological structure of networks, but also how information diffuses on them~\cite{lopez2008diffusion, rodriguez2011uncovering, xu2010information, harush2017dynamic}. For example, in social networks, we may be interested in understanding viral outbreaks in a population, or breaking news spreads in an online setting.

The most fundamental assumption of network science and machine learning applied to graphs is that the network structure begets the function of the networked system. First discussed in abstract terms by Georg Simmel in the 1890s~\cite{simmel} and in the language of graph theory by Jacob Moreno and Helen Jennings in the 1930s~\cite{moreno_jennings}, this assumption is close to the core structuralism---a pillar of 20th-century (primarily social) science. It suggests that we can infer the function of nodes from their position in the network.

The foundational functional concept is a node's importance. But to operationalize a concept like importance, we must consider many specifics about the system. Some questions may include: What is the objective of the system? What dynamics operate on it? Or what are the possible interventions? \emph{Influence maximization (IM)}\footnote{Throughout this work, we use the terms {\it influenced}, {\it activated} and {\it infected} interchangeably when referring to a node's state.} assumes a scenario where some \emph{diffusion} (in the mathematical and physical literature, also known as \emph{spreading}) process can happen on the network and we want this process to reach as many nodes as possible. This diffusion process is triggered by some seed nodes and the IM problem is to identify the seed nodes that maximize the number of nodes affected by the diffusion.

Viral (or word-of-mouth) marketing is an obvious potential application area that fits the assumptions above~\cite{domingos2001mining, leskovec2007dynamics, hinz2011seeding, lu2016vital, bhattacharya2019viral}. Recommendation systems are another relevant area of application~\cite{herlocker2004evaluating, bobadilla2013recommender, aggarwal2016recommender, zhang2021artificial, huang2022influence}, as is seeding public health campaigns~\cite{yadav2016using, yadav2018bridging, wilder2017uncharted, wilder2018end}. However, by analogy to network centrality---another family of conceptualizations under the umbrella term of importance---influence maximization is interesting for a wider area of problems. Protecting critical infrastructure~\cite{liu2020review} or safeguarding against bioterrorism~\cite{WANIEK2022104956} could also benefit from influence maximization studies. It is also a possible approximation~\cite{holme_three_faces} for distinct but related scenarios like the vaccination problem~\cite{holme_vacci,holme_vaccine} (finding nodes whose removal would hinder a diffusion event as effectively as possible), and sentinel surveillance~\cite{christakis_fowler,holme_sentinel} (identifying nodes that would be suitable probes for early and reliable detection of diffusion events).

\begin{figure*}[!t]
    \centering
    \includegraphics[width=0.9\textwidth]{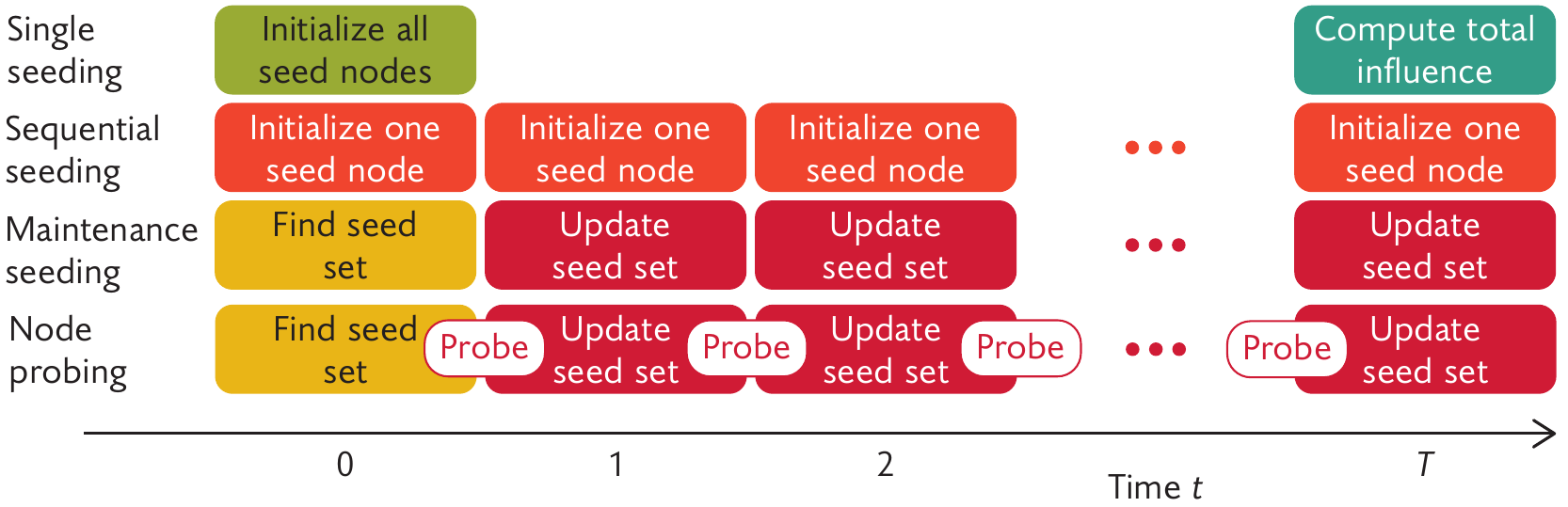}
    \caption{Different temporal influence maximization paradigms compared in this work.}
    \label{fig:overview}
\end{figure*}

Kempe et al.~\cite{kempe2003maximizing} first formulated the IM problem in their seminal work and ever since, the problem has been explored extensively in the computer science, statistical physics and information science literature. The majority of work considers static networks where the nodes and edges are fixed~\cite{kempe2003maximizing, bharathi2007competitive, chen2009efficient, goyal2011celf++}. Most methods either efficiently emulate the influence spread function, or use some heuristic to rank nodes by importance. We eschew extended discussion of the static IM problem and instead refer interested readers to reviews in~\cite{li2018influence, azaouzi2021new}. In many real-world situations, the static assumption is violated as networks have temporal variation with links forming and disappearing~\cite{holme2012temporal}. The topic of this review is IM techniques and analyses for such dynamically varying networks. For example, on a social media platform like Meta's Facebook or Twitter, users are constantly joining the platform in addition to updating their connections. Thus, to achieve satisfactory influence spread, researchers must account for these dynamics.

There are several key challenges associated with the IM problem. First, simply calculating the expected influence spread is \#P-hard for many models~\cite{goyal2011celf++}. One common approach to circumvent this issue is Monte Carlo (MC) simulations, where the diffusion process is simulated a large number of times and the average number of influence nodes is used to estimate the influence spread~\cite{kempe2003maximizing, ohsaka2014fast}.  Still, selecting the optimal seed nodes is NP-hard~\cite{kempe2003maximizing}. Therefore, many researchers employ heuristics so finding the globally optimal solution is rarely guaranteed. In temporal networks, there is an additional challenge stemming from the interplay of dynamic variation in edge sets and diffusion processes.

In this work, we provide a comprehensive review of the existing literature on IM on temporal networks and elucidate important future research areas. One main contribution of this review is our keen eye towards the challenges associated with deploying these methods in practice. While there has been significant research on the static and temporal IM problem, we found that there is minimal research on using these methods ``in the field.'' Thus, we highlight the utility of each method for practitioners. This differentiates the present work from reviews in~\cite{yang2019influence} and~\cite{hafiene2020influential}. Additionally,~\cite{yang2019influence} focuses on influence analysis rather than strictly influence maximization and we also discuss several tasks not mentioned by the authors, e.g., sequential seeding and the {\it ex ante} setting.

There are four main sections of this review. For the remainder of this section, we introduce the necessary prerequisites for studying the IM problem. In Section \ref{sec:one}, we discuss ``single seeding'' methods which select a single seed set at the beginning of the diffusion process. This problem is the natural extension of the static IM problem. We classify the existing methods into three categories while also discussing methods which analyze the diffusion process. The topic of Section \ref{sec:mult} is methods which {\it repeatedly} choose seed nodes as the network evolves. Within this category, some methods activate nodes at different times throughout the diffusion process while others ``maintain'' an influential seed set. Additionally, we consider the node probing problem where the future evolution of the network can only be known by probing a small subset of nodes and this partially visible network is used for IM seeding. See Figure \ref{fig:overview} for an overview comparing these different paradigms. Real-world implementation of IM algorithms is the topic of Section \ref{sec:real} where we primarily focus on the problem of increasing HIV awareness amongst homeless youth. This application highlights the many challenges associated with deploying IM algorithms. Finally, we conclude in Section \ref{sec:conc} with important areas for future research, including the {\it ex ante} setting, model misspecification, and the temporal relationship between the diffusion process and network evolution. Throughout the paper, we give special attention to the functionality of these methods on real-world problems.

\begin{figure*}[!t]
    \centering
    \includegraphics[width=0.9\textwidth]{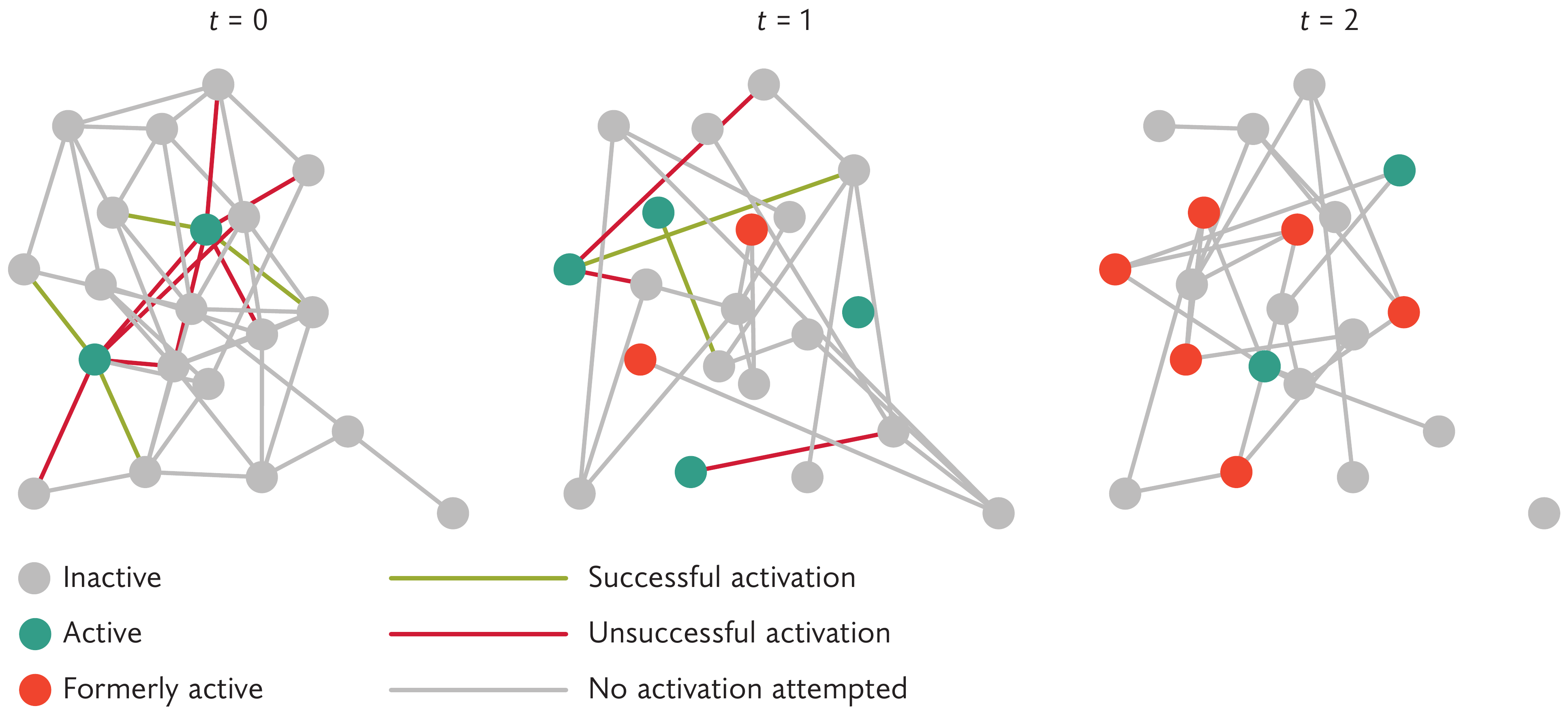}
    \caption{Example of influence diffusion on a toy temporal network using the independent cascade (IC) model. Activated nodes (green) try to influence all of their neighbors before becoming inactive (red) in the next time step. The number of nodes currently or formerly activated corresponds to the total influence spread.}
    \label{fig:ic}
\end{figure*}

\subsection{Notation}
We begin by defining common notations used throughout the paper. Let $\mathcal G=(G_0,\dots,G_{T-1})$ be a temporally evolving network over $T$ time stamps. Typically the graph snapshots $G_t$ occur over evenly spaced time intervals, i.e., $t_k-t_l$ is constant for all $k,l$. For each $t$, let $G_t=(V_t,E_t)$, where $V_t$ is the set of vertices and $E_t$ is the set of edges. Typically, $V_t\equiv V$ and does not vary with time. Let $n=|\cup_t V_t|$ and $m=\sum_t |E_t|$ be the total number of nodes and edges in the network, respectively. Additionally, let $A_{ij}(t)$ denote the corresponding adjacency matrix for graph $G_t$ where $A_{ij}(t)=1$ if there is an edge from node $j$ to node $i$ at time $t$, and 0 otherwise. In an undirected network, $A_{ij}=A_{ji}$ for all $i,j$, while $A$ may be asymmetric for a directed network. Let $N_i(t)$ be the set of incoming neighbors of node $i$ at time $t$, i.e., $N_i(t)=\{j:A_{ij}(t)=1\}$.

\subsection{Diffusion mechanisms}

In order to study IM, it is necessary to describe the diffusion of influence on a network. The most common diffusion models are the independent cascade (IC), linear threshold (LT) and susceptible-infected-recover (SIR) models. In the IC model~\cite{kempe2003maximizing, saito2008prediction, shakarian2015independent}, influenced nodes have a single chance to activate their uninfluenced neighbors. Specifically, let $p_{ij}$ be the probability that node $i$ influences nodes $j$. If node $i$ is infected at time $t-1$, then node $j$ becomes infected at time $t$ with probability $p_{ij}$, assuming $A_{ij}(t-1)=1$. From time $t$, node $i$ can no longer influence its neighbors. Then the total influence spread is the number of nodes that were active at any point. 

In Figure \ref{fig:ic}, we show the influence diffusion process on a toy network using the IC model. Two nodes (green) are initially selected for the seed set and begin in the active state. These nodes attempt to activate all of their neighbors, but only some attempts are successful (green and red edges are successful and unsuccessful activations, respectively). In the next time step, the newly activated nodes now attempt to influence all of their current neighbors, while the previously activated nodes become inactive. This process continues one more time step, and the number of nodes in the active (green) or formerly active (red) state is the total influence spread for this seed set (eight nodes). Due to the stochastic nature of the process, even if the process is repeated with the same seed nodes, the total diffusion spread may differ.

The SIR model~\cite{pastor2015epidemic, erkol2022effective} is similar to the IC model, but now each activated node has a fixed probability $\lambda$ of infecting its unactivated neighbors. Moreover, a node can activate its neighbors as long as it is in the infected state. Each infected node, however, has probability $\mu$ of ``recovering'' and being unable to activate its neighbors. The total influence spread is the final number of nodes in the infected and recovered state. If $p_{ij}=\lambda$ for all $i,j$ and $\mu=1$, then the SIR and IC models are equivalent. Additionally, the SIR model reduces to the susceptible-infected (SI) model~\cite{osawa2015selecting, murata2018extended} if $\mu=0$.

In the LT model~\cite{kempe2003maximizing, chen2010scalable, pathak2010generalized, shakarian2015independent}, each node is randomly assigned a threshold $\theta_i$ and each edge endowed with a weight $b_{ij}$. If the sum of weights for a node's infected neighbors exceeds its threshold, then this node becomes infected, i.e., node $i$ is activated if $\sum b_{ij}>\theta_i$ where the sum is over all infected neighbors of $i$. 

\subsection{Problem statement}

With notation and diffusion mechanisms in hand, we formally define the IM problem. Let $\mathcal G$ be some dynamic network and let $\mathcal D$ be a diffusion mechanism, e.g., IC or SIR. We define $\sigma(S)$ as the expected number of influenced nodes for seed set $S$ and for diffusion process $\mathcal D$ on graph $\mathcal G$. We suppress the dependence of $\mathcal G$ and $\mathcal D$ on $\sigma(\cdot)$, but stress that its behavior is highly dependent on both. For a fixed $k=|S|$, we seek the seed set $S$ which maximizes $\sigma(S)$, i.e.,
\begin{equation}\label{eq:prob}
    S^*
    =\arg\max_{S\subseteq V,|S|=k}\sigma(S).
\end{equation}
Perhaps the simplest approach to approximate (\ref{eq:prob}) is evaluating $\sigma(S)$ via MC simulations and choosing the node which marginally leads to the largest gain in influence spread, as outlined in Algorithm \ref{alg:greedy}. In the static setting, this greedy algorithm provably yields a result within a factor of $(1-1/e)$ of the global optimum~\cite{kempe2003maximizing}. MC simulations are computationally intensive, however, so many methods focus on efficiently computing the influence spread before employing a greedy algorithm.

\begin{algorithm}[]
\SetAlgoLined
\KwResult{Seed set $S$}
 {\bf Input: } Temporal network $\mathcal G$, seed size $k$\;
 
$S= \varnothing$\;
 
 \For{$i \text{ in } 1:k$}{

 $v = \arg\max_{u\in V\setminus S}\{\sigma(S\cup\{u\}) - \sigma(S)\}$\;  
 
 $S = S\cup \{v\}$\;
 
  }
 \caption{Greedy influence maximization}
 \label{alg:greedy}
\end{algorithm}

Another common paradigm for selecting seed nodes that avoids direct calculation of the influence function is based on node ranking. Nodes are ranked based on some measure of importance, e.g., degree or centrality, and the $k$ nodes with the largest value are chosen for the seed set. While much faster than greedy algorithms, these approaches yield no theoretical guarantees and may choose nodes that ``overlap'' their influence. For example, if two nodes have high degrees but share many common neighbors, then seeding both nodes may not be optimal as their influence will spread to the same nodes.

{\it Reverse Reachable (RR) sketches} also choose seed nodes without direct computation of $\sigma(S)$. For a given time $t$ and for each edge $(i,j)$, we randomly draw $Z_{ij}\sim\mathsf{Bernoulli}(p_{ij})$ where $p_{ij}$ is the probability that node $i$ influences node $j$, and keep the subgraph with $Z_{ij}=1$. These edges are sometimes referred to as the ``live'' or ``active'' edges. Once this subgraph is constructed, the source and destination of each (directed) edge are reversed before randomly selecting a node. Finally, a breadth-first search is conducted from this randomly selected node and all nodes reached by this search are kept for this particular RR-sketch. Essentially, the nodes in this set are those that can influence the selected node through the diffusion process. This process is repeated a large number of times to yield a set of RR-sketches.

\section{Single seeding}\label{sec:one}
The classical IM problem is where a practitioner selects a set of seed nodes at time $t=0$ in order to maximize the influence spread at time $t=T$. Most solutions either estimate the influence spread via probabilities, or use some heuristic to rank nodes by influence. For the majority of these methods, the complete temporal evolution of the network is assumed to be known, but some relax this assumption. We also discuss several works which do not present novel algorithms, but rather analyze existing methods and/or diffusion processes.
\subsection{Algorithms}
First, we discuss algorithms which solve the single seeding temporal IM problem.

\subsubsection{Greedy}
The greedy algorithm of Kempe et al.~\cite{kempe2003maximizing} naturally extends to the temporal setting: nodes with the largest marginal gain in influence spread are added to the seed set incrementally as in Algorithm \ref{alg:greedy}. The only difference from~\cite{kempe2003maximizing} is that the expected influence spread is computed on a temporally evolving network. This method is considered the ``gold standard'' of IM algorithms and can easily be adapted to any diffusion model. On the other hand, this algorithm suffers from a high computational cost due to the repeated MC simulations required for computing the influence spread.

\subsubsection{Probability of influence spread}
Since the costly step of the greedy algorithm is computing the influence spread, several heuristics exist which use the probability of a node's activation in order to approximate $\sigma(S)$. In Aggarwal et al.~\cite{aggarwal2012influential}, $\pi_i(t)$ is the probability that node $i$ is activated at time $t$. Assuming the network is a tree, if $p_{ji}(t)$ is the probability that node $j$ activates node $i$ at time $t$, then the probability that node $i$ is activated at time $t+1$ is
\begin{equation}\label{eq:agg}
    \pi_i(t+1)
    =\pi_i(t) + (1-\pi_i(t))\times \left(1-\prod_{j\in N_i(t)}(1-\pi_j(t)p_{ji}(t))\right)   
\end{equation}
where $N_{i}(t)$ is the set of incoming neighbors of node $i$ at time $t$. The first term is the probability that the node was already activated during the previous time step while the second term is the probability that it was not previously activated but becomes so in the current step. The authors initialize $\pi_i(1)=1$ if $i\in S$ and 0 otherwise. For each $i$, $\pi_i(t)$ iteratively updates via (\ref{eq:agg}) and $\sum_i \pi_i(T)$ estimates $\sigma(S)$. Aggarwal et al.\ use this procedure and a greedy algorithm to choose the seed set. The authors assume that $p_{ij}(t)$ is an increasing function of the length of time that an edge between nodes $i$ and $j$ persists. Additionally, this method eschews the standard, equally-spaced graph snapshots in favor of times corresponding to structural changes based on the number of edge updates. The approach also can find the most likely seed nodes for a given diffusion pattern.

Osawa and Murata~\cite{osawa2015selecting} take an analogous approach to Aggarwal et al.~but use the SI model for diffusion. In fact, this method is equivalent to~\cite{aggarwal2012influential} if $p_{ij}(t)=\lambda$ for all $i,j$ and times $t$. Osawa and Murata show this approach slightly overestimates the true influence spread and prove that the associated greedy algorithm's computational complexity is $O(nmk)$. In simulations, this method outperforms broadcast~\cite{grindrod2011communicability} and closeness centrality~\cite{holme2012temporal} heuristics. It also yields comparable performance to a standard greedy algorithm but is two orders of magnitude faster. Additionally, Osawa and Murata show that centrality heuristics perform worse on networks with strong community structure due to nodes' overlapping influence.

Erkol et al.~\cite{erkol2020influence} extend this paradigm to the SIR model. If $\pi_i(t)$ is defined as above and $\rho_i(t)$ is the probability that node $i$ is in state R at time $t$ (such that $1-\pi_i(t)-\rho_i(t)$ is the probability of being in state $S$), then the SIR dynamics are defined by
\begin{align}\label{eq:sir1}
    \pi_i(t) &= (1-\mu)\pi_i(t-1) +(1-\pi_i(t-1)-\rho_i(t-1))\\\notag
    &\times \left(1-\prod_{j\in N_i(t)}(1-\lambda \pi_j(t-1))\right)\\\label{eq:sir2}
    \rho_i(t) &=\rho_i(t-1)+\mu \pi_i(t-1).
\end{align}
A greedy algorithm chooses the seed nodes based on the influence spread estimate $\sum_i \{\pi_i(T) + \rho_i(T)\}$. Erkol et al.~study the performance of the method when the network is noisy or incomplete, i.e., the temporal snapshots are randomly re-ordered, only the first snapshot is available, and the network is aggregated into a single snapshot. The authors find that the order of snapshots is crucial and ignorance about $G_1$ causes the algorithm to suffer. Indeed, in many cases, knowing only $G_1$ is sufficient for large influence spread, while the aggregation approach consistently performs poorly. Erkol et al.~also show that if $\mu$ is large, then central nodes in the first few layers are the best influence spreaders, whereas for small $\mu$, nodes must be central in many layers to make for optimal seed nodes.

\subsubsection{Node ranking heuristics}
Rather than estimate the influence spread, the following methods rank nodes by influence and select the top $k$ as seed nodes. The earliest approach comes from Michalski et al.~\cite{michalski2014seed}. The authors adopt the LT model and assume the first $T/2$ snapshots are available to select seed nodes, but the diffusion process occurs on $G_{T/2},\dots, G_{T-1}$. Some static measure of node importance $m_i(t)$ is computed for all nodes $i$ and snapshots $G_t$. The values across $t$ are combined by down-weighting older values to yield a single metric $\theta_{i}$ for each node, i.e.,
\begin{equation}
    \theta_{i}
    =\sum_{t=0}^{T/2-1} f(m_i(t), t)
\end{equation}
where $f(x,t)$ is an increasing function in $t$. The $k$ nodes with largest $\theta_i$ for a given $f(\cdot)$ are chosen as seed nodes. Michalski et al.~find that the following combinations of metrics $m_i(t)$ and forgetting mechanisms $f(\cdot)$ yield the largest influence spread: out-degree and in-degree with exponential forgetting $(f_{exp}(x,t)=e^{-t}x)$\footnote{This was the forgetting mechanism reported in the original paper but there appears to be some error since this function clearly {\it decreases} with $t$.}, total degree and logarithmic forgetting $(f_{log}(x,t)=\log_{T/2-1-t+1}(x))$, betweenness centrality and hyperbolic forgetting $(f_{hyp}(x,t)=(T/2-1-t-1)^{-1}x)$; and closeness centrality with power forgetting $(f_{pow}(x,t) = x^t)$. The authors also vary the number of aggregated snapshots used to compute the metrics and find that the finest granularity performs the best. Indeed, treating the network as static by aggregating $G_0,\dots,G_{T/2-1}$ into a single graph yields the lowest influence spread on $G_{T/2},\dots,G_{T-1}$, thus demonstrating the importance of accounting for the temporal variation in the network.

Another node ranking heuristic comes from Murata and Koga~\cite{murata2018extended}. Using the SI model, the authors extend several static measures of importance to the temporal settings. In particular, the dynamic degree discount algorithm extends~\cite{chen2009efficient}. First, the node with largest dynamic degree $D_T(v)$ is added to the seed set, where
\begin{equation}
    D_T(v)
    =\sum_{t=2}^T \frac{|N_{v}(t-1)\setminus N_{v}(t)|}{|N_{v}(t-1)\cup N_{v}(t)|}|N_{v}(t)|
\end{equation}
and $N_v(t)$ is the neighbors of node $v$ at time $t$. Once a node is selected, the value of the dynamic degree for its neighboring nodes is decreased and the process repeats until $k$ nodes are selected. The authors show that the complexity of this method is $O(k\log n + m + mT/n)$ but that it only is valid for the SI model. Murata and Koga also propose Dynamic CI as an extension of~\cite{morone2015influence} based on optimal percolation and Dynamic RIS as an extension of~\cite{borgs2014maximizing, tang2014influence} based on RR sketches. In simulations, all methods perform comparably to Osawa and Murata~\cite{osawa2015selecting} but are significantly faster. The authors also show that when $\lambda$ is large, choosing the optimal seed nodes is less important as many seed sets yield comparable influence spread.

Recently,~\cite{michalski2020entropy} propose another node-ranking heuristic. The authors postulate that, for the IC model, nodes with large variability in their neighbors should be chosen to maximize spread. They quantify neighborhood variability with an entropy measure that rewards nodes for changing their neighbors in subsequent graph snapshots and the $k$ nodes with the largest value are chosen for the seed set. The measure is computed on the first $T/2$ snapshots while the influence is calculated on the second half of the graph's evolution, similar to Michalski et al.~\cite{michalski2014seed}. The authors note, however, that this metric may not make sense for the LT model which requires the number of activated neighbors to ``build up'' for a node to become infected. As the method depends on the neighborhood set of each node, its complexity is $O(m)$.

To summarize the methods in the previous two subsections, we include Table \ref{tab:comp} which compares them across several metrics.

\begin{table*}[!t]
    \centering
    \begin{tabular}{l|cccccc}
        Method & Paradigm & Model & Complexity \\\hline
        Greedy & Greedy & Any & $O(nmkR)$\\
        Aggarwal et al.~\cite{aggarwal2012influential} & Approx. & ICt & $O(nmk)$\\
        Osawa and Murata~\cite{osawa2015selecting} & Approx. & SI & $O(nmk)$\\
        Erkol et al.~\cite{erkol2020influence} & Approx. & SIR & $O(nmk)$\\
        InExp~\cite{michalski2014seed} & Node rank & LT & $O(m)$\\
        OutExp~\cite{michalski2014seed} & Node rank & LT & $O(m)$\\
        TotalLog~\cite{michalski2014seed} & Node rank & LT & $O(m)$\\
        BetHyp~\cite{michalski2014seed} & Node rank & LT & $O(nm+n^2T)$\\
        CloPow~\cite{michalski2014seed} & Node rank & LT & $O(nm+n^2T)$\\
        Dynamic Degree~\cite{murata2018extended} & Node rank & SI &$O(k\log n + m +mT/n)$ \\
        Dynamic CI~\cite{murata2018extended} & Node rank & SI& $O(n\log n +mT/n)$\\
        Dynamic RIS~\cite{murata2018extended} & Node rank & SI &$O(\theta dkl^2(m+n)\log^2n/\epsilon^3)$\\
        Entropy~\cite{michalski2020entropy} & Node rank & IC & $O(m)$
    \end{tabular}
    \caption{Comparison of different algorithms for single seeding temporal influence maximization. Paradigm: {\it Approx}: estimates the probability that a node is activated. {\it Node rank}: uses a node ranking heuristic. Model: Diffusion model. Note that ICt refers to~\cite{aggarwal2012influential}'s model which does not neatly line up into any of the standard models but it similar to an IC model with time component. Complexity: number of flops to implement the algorithm. $R$ is number of MC simulations; $\theta,d,l$ are tuning parameters in the Dynamic RIS algorithm.} 
    \label{tab:comp}
\end{table*}

\subsection{Analysis}
We turn our attention to methods that do not propose a novel IM algorithm, but rather analyze the existing algorithms and/or diffusion mechanisms. In order to better model information propagation, Hao et al.~\cite{hao2011influence} propose two novel diffusion models where a node's propensity of activation depends on the number of past attempts to activate it. In the time-dependent comprehensive cascade model, an active node still only has a single chance to activate its neighbors, but the probability of being infected can either increase, decrease or be unaffected by the number of previous attempts on that node. The authors also propose a dynamic LT model where the node's activation threshold depends on the number of previous activation attempts. Hao et al.~proposes a time series-based approach to empirically determine the effect of past activation attempts on infection probabilities.

Gayraud et al.~\cite{gayraud2015diffusion} study the behavior of the influence spread function under several novel diffusion models while also allowing seed nodes to be activated at different times. Let $f:2^V\to\mathbb R$ be a set function. If $S\subseteq V$, then $f$ is {\it monotone} if $f(S\cup\{v\})-f(S)\geq 0$ for all $v\in V\setminus S$ and $S\subseteq V$. It is {\it submodular} if $f(A\cup\{v\})-f(A)\geq f(B\cup\{v\})-f(B)$ for $A\subseteq B$ and $v\in V\setminus B$. In other words, the monotone property implies that adding a node never decreases the influence spread and submodularity means that there is diminishing returns for adding more nodes. Additionally, the authors define a seeding strategy to be {\it timing insensitive} if all nodes should be activated at time $t=0$ and {\it timing sensitive} otherwise. In the transient evolving IC model (tEIC), infected nodes at time $t-1$ have one chance to infect their neighbors at time $t$. The authors prove that this diffusion mechanism is neither monotone nor submodular. In contrast to the tEIC model, the persistent EIC model assumes that a node tries to activate its neighbors the first time that the two nodes have a link. If the activation probabilities are constant in time, Gayraud et al.~prove that this model is monotone, submodular and timing insensitive. If the activation probabilities dynamically vary, then the influence function is neither monotone nor submodular and is timing-sensitive. The authors propose similar extensions to the LT model. The transient ELT model only considers weights from active neighbors at the current snapshot whereas the persistent ELT model sums all weights from neighbors activated during any previous time. These models are monotone, not submodular and timing insensitive, and monotone, submodular and timing insensitive, respectively. The key contribution of this paper is that if a model is timing-sensitive, the seed nodes should be activated throughout the diffusion process, as opposed to all at $t=0$. The authors also show that choosing seed nodes based on aggregating all graph snapshots does not perform well for any model.

\begin{figure}
    \centering
    \includegraphics[scale=0.35]{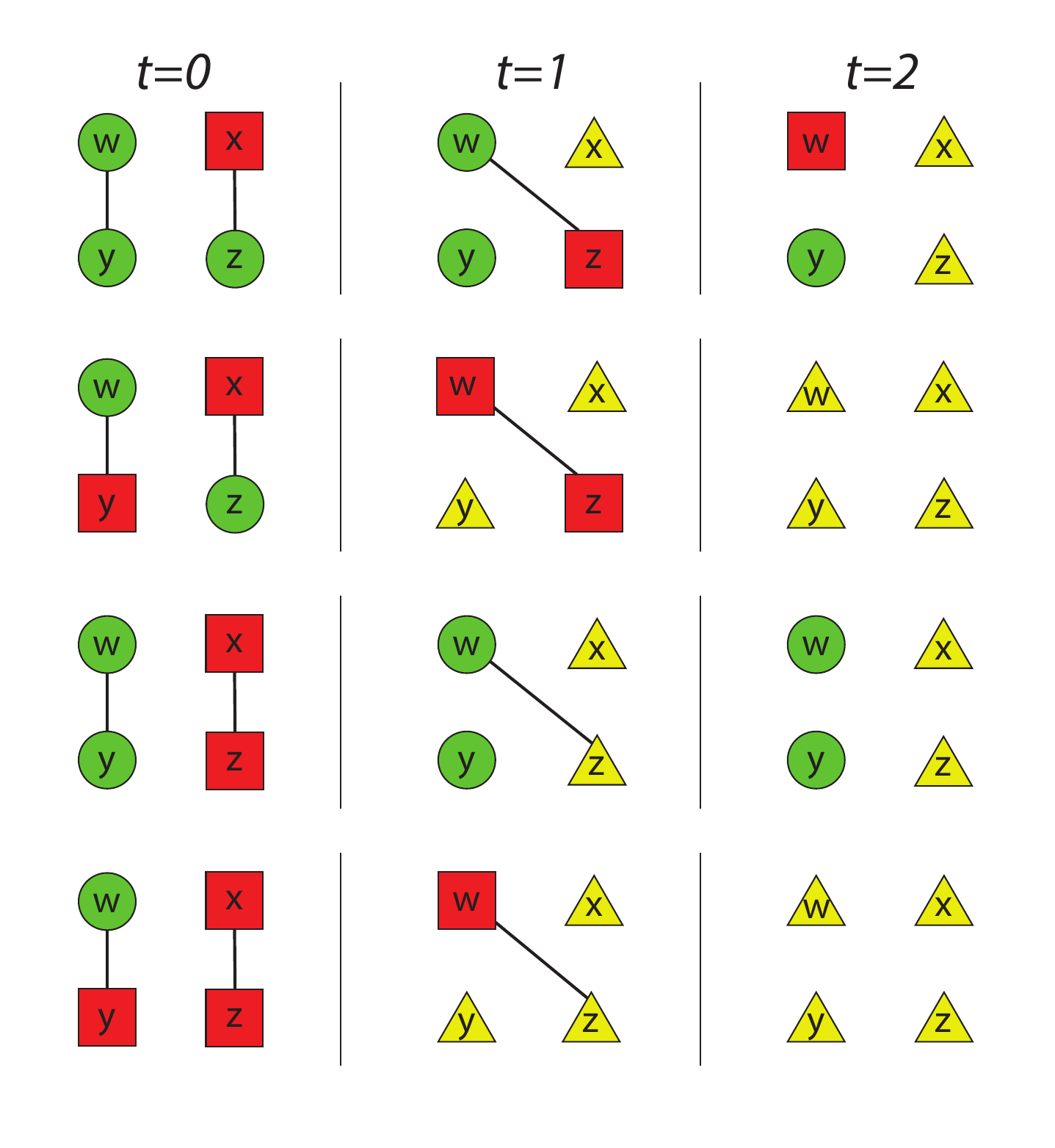}
    \caption{Toy example showing the violation of the sub-modularity property in the temporal SIR model. Green circles are susceptible nodes, red squares are infected nodes, and yellow triangles are recovered nodes. Each row represents the diffusion process for a particular seed set with $\lambda=\mu=1$. In the first row, $S_1=\{x\}$ and the total influence spread is three nodes. In row three, $S_2=\{x,z\}$, but the total influence spread is only two nodes. Thus, $S_1\subset S_2$ while $\sigma(S_1)>\sigma(S_2)$. Reproduced from \cite{erkol2022effective} with permission of the author.}
    \label{fig:sub}
\end{figure}

The submodularity of the influence function is also studied in~\cite{erkol2022effective}, this time under the SIR model. The authors show that if $\mu=0$ (SI model), the influence function is submodular, but loses this property when $\mu>0$. Effectively, the violations come from nodes in state R ``blocking'' paths to nodes in state S, as demonstrated by a toy example in Figure \ref{fig:sub} (reproduced with permission of the author). 
A relaxation of the submodularity property, $\gamma$-weakly submodular, is also not achieved in the SIR model. A function $f$ is $\gamma$-weakly submodular if for $A\cap B=\varnothing$ and $0<\gamma\leq 1$,
\begin{equation}
    \sum_{v\in B}f(A\cup\{v\})
    \geq \min\left\{ \gamma f(A\cup B), \frac1\gamma f(A\cup B) \right\}.  
\end{equation}
The authors then empirically check the number of violations of the submodular criteria in real networks. They find that if nodes are randomly selected, the criteria is frequently violated. If nodes are selected based on a greedy algorithm, however, the submodularity property is rarely violated. Thus, the influence function is effectively submodular. Now, since the influence function is not submodular, there is no theoretical guarantee that the greedy algorithm adequately approximates the optimal solution. In spite of this, compared with a brute-force algorithm on real-world networks, the greedy algorithm still yields results within 97\% of the optimal solution.

Lastly, although not pertaining explicitly to IM, we briefly discuss~\cite{albano2013matter}. This work studies the relationship between graph topological evolution and diffusion processes by analyzing which part of the diffusion is owed to the diffusion mechanism, and which to graph dynamics. The authors consider two timing mechanisms: {\it extrinsic} time based on seconds between interactions, and {\it intrinsic} time based on changes or transitions in the network. While researchers typically use extrinsic time, the authors argue that intrinsic time may be more sensible in many cases. Using the SI model and intrinsic time, the observed diffusion is governed more by the diffusion mechanism than the evolution of the network. Using extrinsic time, conversely, the topological changes in the network greatly affect the diffusion. Thus, the diffusion process is highly dependent on the timing method.

\subsection{Discussion}
In the previous subsections, we presented the leading methods for choosing a single seed set in temporal IM. Each method either estimates the influence spread, or ranks nodes based on a heuristic. Aggarwal et al.~\cite{aggarwal2012influential}, Osawa and Murata~\cite{osawa2015selecting} and Erkol et al.~\cite{erkol2020influence} proposed analogous approaches with the only difference being in the diffusion model. These methods maintain many of the desired properties of the greedy algorithm, but are computationally less intensive. The node ranking metrics of~\cite{michalski2014seed},~\cite{murata2018extended} and~\cite{michalski2020entropy} are even faster since they avoid the costly influence spread calculation. In theory, these methods are not guaranteed to perform as well as a greedy-based algorithm, but~\cite{murata2018extended}, for example, shows they still maintain good performance.   

A key challenge in implementing these methods on real-world problems is the requirement that the entire topology of the network be known. Save~\cite{michalski2014seed, michalski2020entropy}, each method assumes that the evolution of the network $G_0,\dots,G_{T-1}$ is known at time $t=0$ when the seed nodes are selected. Of course, in practice, it is unreasonable for a practitioner to know the future topology of the network, so it is not obvious how to apply these methods in this case. To address this issue,~\cite{yanchenko2023link} propose a link prediction approach for {\it ex ante} temporal IM. Using the SI model, the authors use the first $p$ snapshots to train a link prediction algorithm and then predict the network topology for $G_{p},\dots, G_{T-1}$. An existing temporal IM algorithm is applied to these predicted networks to choose the seed sets. In many cases, finding seed nodes on a simple aggregation of $G_0,\dots,G_{p-1}$ performs as well as the more complicated link prediction methods. This finding is at odds with Michalski et al.~\cite{michalski2014seed} and Erkol et al.~\cite{erkol2020influence} who showed poor performance of IM algorithms on aggregated networks. These papers, however, assumed different diffusion mechanisms, so it is possible that aggregating only works well for the SI model. Another practical consideration for applied researchers is the size of the network. If working with a relatively small social network, then a greedy algorithm is reasonable, whereas a node ranking heuristic is mandatory for large online social networks with millions of nodes. Finally, the diffusion mechanism must be carefully chosen based on the application's domain, as certain methods are only applicable to specific mechanisms.

\section{Multiple seedings}\label{sec:mult}
In the previous section, we considered IM algorithms for temporal networks where all seed nodes are activated at time $t=0$. Now we discuss methods where nodes are seeded at different points throughout the evolution of the network, or where the seed set is updated at each time step.

\subsection{Sequential seedings}

Related to the single seeding problem, consider a single seed set $S$, but instead of activating all nodes at $t=0$, nodes activate {\it sequentially} as the network evolves. This problem involves not only choosing {\it which} nodes to include in the seed set, but also {\it when} to activate them. 
%The earliest work on a variant of this problem comes from~\cite{li2012finding}. While working with static networks, the authors consider the scenario where, given a seed set $S$, $S'\subseteq S$ drop out of the network and the goal is to replace these nodes. Nodes with high degrees and edges in many communities are shown to be the best choices to fill in $S$.

\begin{figure}
    \centering
    \includegraphics[scale=0.50]{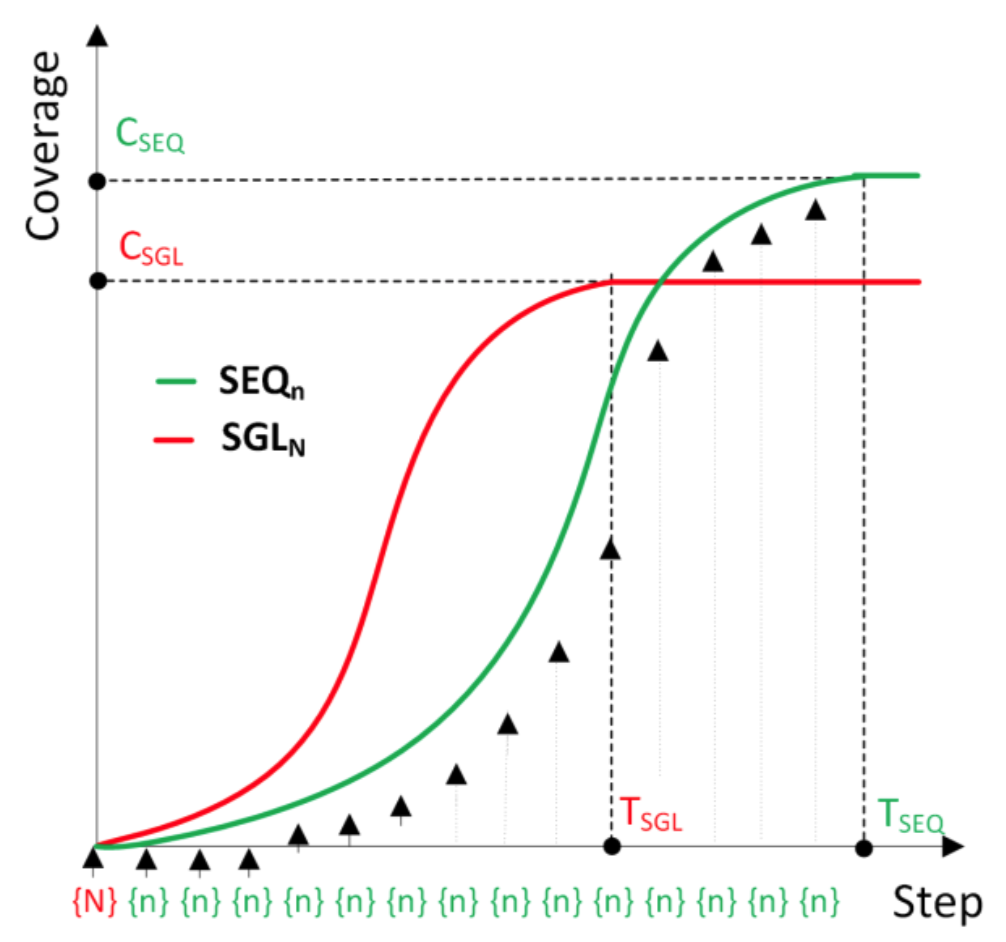}
    \caption{Comparison of the influence spread for the sequential seeding strategy with that of single seeding. The red curve activates all nodes at $t=0$ while the green curve activates nodes one at a time. Single seeding leads to greater influence spread at first, but the sequential seeding strategy ultimately leads to more spreading. Reproduced from \cite{michalski2020effective} with permission of the author.}
    \label{fig:seq}
\end{figure}

Michalski et al.~\cite{michalski2020effective} focus on the seed activation step of this problem. The authors consider a variant of the IC model where a single node is activated and the diffusion occurs until no more activations are possible. Then the next node is activated and the process continues. In this setting, Michalski et al.~use a simple seed selection method based on degrees. First, the node with the largest degree is activated. Once the diffusion process finishes, the uninfected node with largest degree is activated and the process continues until $k$ nodes have been seeded. This method is compared with activating the $k$ nodes with largest degree at time $t=0$. When $t$ is small, activating all nodes at once leads to a larger influence spread, but as $t$ increases, the sequential seeding strategy outperforms the single seeding, as shown in Figure \ref{fig:seq} (reproduced with permission of the author).

Tong et al.~\cite{tong2016adaptive} consider another variation of the sequential seeding problem where seed set nodes are unsuccessfully activated with some probability. They propose a greedy algorithm which maximizes the marginal gain in influence spread given the current diffusion. Towards this end, the authors derive a closed-form expression for the expected number of influenced nodes by constructing an auxiliary graph with extra nodes and edges based on possible seed sets and propagation probabilities. The greedy algorithm is shown to yield results within $(1-1/e)$ of the optimal influence spread while the computational burden is mitigated with the Lazy-forward technique~\cite{leskovec2007cost}. Additionally, Tong et al.~prove that the strategy outlined in Michalski et al.~\cite{michalski2020effective} of seeding nodes one at a time and waiting for the diffusion process to finish before activating the next node is the optimal seeding strategy for any temporal graph.

\subsection{Maintenance seeding}
In a highly dynamic network, the optimal seed set may change with time. For example, in a long-term marketing campaign on Twitter, the active users and followers change over the duration of the campaign. Thus, it is necessary to {\it maintain} or {\it update} the seed set $S_t$ such that it provides maximum influence spread on $G_t$ for all $t$. This problem is known as {\it maintenance seeding}. Maintenance seeding is markedly different from static seeding as now $k$ nodes are activated at each time step $t$ in order to maximize the diffusion on $G_t$. Thus, this process is analogous to a sequence of static IM problems.

Chen et al.~\cite{chen2015influential, song2016influential} first study this problem under the name ``influential node tracking.'' The authors assume that the topology of the network is known at the next time step $G_{t+1}$ and use the IC model for diffusion. Using the seed set from the current snapshot $S_t$, Chen et al.~employ an interchange heuristic~\cite{nemhauser1978analysis} to efficiently update the seed set and prove that the solution is guaranteed to be within $1/2$ of the optimal spread. Effectively, this method swaps one node in $S_t$ with one node in $V\setminus S_t$ to maximize the marginal gain in influence spread. Since evaluating the marginal gain for every node in $V\setminus S_t$ is expensive, the authors only consider nodes with the largest marginal gain upper bound. If the upper bound for node $u\in V\setminus S$ is smaller than the marginal gain of another node $v$, then evaluating the influence of node $u$ is unnecessary as its inclusion cannot improve the total influence spread. The proposed algorithm has $O(kn)$ complexity.

Ohsaka et al.~\cite{ohsaka2016dynamic} consider a similar problem for large online networks where nodes and edges are added or removed at each time step. Using the IC model, the authors propose a sketching method akin to RR sets and an efficient data structure to build and store these sketches. A greedy algorithm is then implemented to choose the seed sets. Specifically, the node which is present in the most sketches is chosen as a seed node. Then all sketches which contain that node are removed from consideration, and the node which occurs in the most remaining sketches is chosen for the seed set. This process continues until $k$ nodes are chosen. In addition to the novel data structure, this work proposes heuristics that lead to efficient updates of the sketches at each evolution of the graph, instead of recomputing them from scratch. These heuristics come with theoretical guarantees and lead to algorithmic speed-ups. 

There are several other methods that address this problem. ~\cite{wu2019maximizing} recast it as a bandit problem but tackle it in a similar manner to Ohsaka et al.~by using RR sketches. ~\cite{wang2017incremental} find the optimal seed nodes at $t=0$ and incrementally update them based on investigating parts of the graph which changed significantly between snapshots.~\cite{wang2017real} selects seed nodes based on a sliding window scheme and~\cite{chandran2022dynamic} uses a node's number of triangles to estimate its influence.~\cite{min2020topic} consider a special case by accounting for user attributes in an online social network, including preferred topics of engagement. The authors also account for certain time periods where users may be inactive and allow for a different diffusion model based on the topic. 

Up to this point, each method assumes knowledge of the future topology of the network.~\cite{singh2021link} relax this assumption by predicting the graph structure one time step in the future using a conditionally temporal restricted Boltzmann machine~\cite{li2014deep} and then finding the seed nodes on the predicted graph. The authors use an interchange~\cite{nemhauser1978analysis} heuristic to update the seed set and ideas from~\cite{song2016influential} to improve efficiency. 

Rather than propose a new maintenance seeding algorithm, Peng~\cite{peng2021dynamic} studies the amortized running time, i.e., the amount of time it takes to update the seed nodes at each time step. Even though the current algorithms efficiently update the seed set in $O(n)$ time for each $t$, the author argues that this is still too slow for large networks. Peng then considers two different graph evolution paradigms, both under either the IC or LT model. First is an {\it incremental} model where a network may only add new nodes and edges. Under this model, Peng shows that an $(1-1/e-\epsilon)$ approximation of the optimal solution is possible with probability $1-\delta$ for amortized running time $O(k\epsilon^{-3}\log^3(n/\delta))$, much faster than $O(n)$. Under a fully dynamic model, however, where nodes and edges can be added and deleted, the author proves that a $2^{(-\log n)^{1-o(1)}}$ approximation is impossible without $n^{1-o(1)}$ amortized run time. Thus, there is no possibility of improving the $O(n)$ run time.

\subsection{Node probing}
While previous methods assumed complete knowledge of the future network topology, the {\it node probing} problem assumes that the future graph snapshots are unknown but can be partially observed by probing the neighborhoods of certain nodes. Here, probing a node means observing its edges. Assuming $G_0$ to be known, the goal is to carefully select which nodes to probe in order to have the most information on the topology of the network in order to effectively implement an IM algorithm. This problem may arise in large online social networks where it is infeasible to observe the activities of all users at every time step. Another relevant application is modeling the social connections within a hard-to-reach population, e.g., homeless youth, as there is no straightforward way to observe all the people (nodes) in this network, yet alone the friendships (edges).

This problem was originally formulated by Zhuang et al.~\cite{zhuang2013influence}. For each $t$, the researcher probes $b$ nodes and observes changes in their neighborhoods. Once the nodes are probed, an IM algorithm is implemented on the (incomplete) visible network. Thus, the goal is to find the ideal probing strategy. The authors propose probing nodes that yield the maximum possible change to the solution of the IM problem. Since the authors use the degree discount algorithm \cite{chen2009efficient}, this reduces to finding the nodes with greatest change in their degree. Specifically, let $\beta(v)$ be the maximum difference in the influence spread of optimal seeds chosen before and after probing node $v$. Moreover, let $S$ be the optimal seed nodes at time $t-1$ and let $S_0$ be the $k$ nodes with the largest in-degree on the most up-to-date graph snapshot. Let $t-c_v$ be the last time stamp at which node $v$ was probed. For $\epsilon>0$, if $z_v=\sqrt{-2c_v\log \epsilon}$, $\beta(v)$ is derived as:
\begin{equation}
    \beta(v)
    =\begin{cases}
        \max\{0, \max_{u\notin S} \hat d_{in}(u)-\hat d_{in}(v)+z_v\}, & v\in S_0\\
        \max\{0,\hat d_{in}(v) - \min_{u\in S}\hat d_{in}(u)+z_v\}, & v\not\in S_0
    \end{cases}   
\end{equation}
where $\hat d_{in}(v)$ is the in-degree of node $v$ based on the most recently probed network. Then node $v^*=\arg\max_{v\in V}\beta(v)$ is probed and the network topology is updated. Once $b$ nodes have been probed, the degree discount algorithm is applied to determine the optimal seed nodes for influence spread.

Han et al.~\cite{han2017influence} study the same problem but focus on communities with high variation as opposed to nodes. The authors postulate that the total in-degree for a community should be relatively stable with time, so if this changes greatly, there must have been a significant change in this community and it is worth probing. The authors use the community detection algorithm of~\cite{zhou2009graph}, and once the community with high variability has been identified, they employ a probing algorithm similar to that of Zhuang et al.~\cite{zhuang2013influence}.

\subsection{Discussion}
We close this section by highlighting important considerations for practical implementation of these methods. In the sequential seeding setting, it is important for researchers to consider how long they can allow the diffusion to take place since static seeding is preferable for small $T$ and sequential for large $T$.  Michalski et al.~\cite{michalski2020effective} also emphasize that the sequential strategy is better suited for independently activated models, e.g., IC and SI, rather than threshold-based models, e.g., LT, so the diffusion model is another important consideration. It would also be interesting to compare static and sequential seeding for more complicated IM algorithms. It is well-known that seeding the top $k$ degree nodes is a relatively poor IM algorithm, so it is unclear whether sequential seeding would perform so much better when combined with different IM algorithms.

For maintenance seeding, we observe that the seeding budget is effectively $kT$ rather than $k$, since $k$ nodes are activated at $T$ different time steps. If $T$ is large, then it may be prohibitive to keep activating $k$ nodes each round. Additionally, this setting implicitly assumes that nodes can be reinfected at successive snapshots, i.e., $S_t\cap S_{t+1}\neq \varnothing$. This may be reasonable in epidemiological settings, for example, where a person can be reinfected by a disease. For marketing campaigns, on the other hand, it is unlikely that a user targeted with an ad in multiple time steps can be expected to have significant diffusion in each case. Thus, the number of times that a user has been infected and this effect on the diffusion mechanism should be considered carefully. Moreover, save the IC model, if a node is infected at time $t$, then it could continue to attempt to infect its neighbors at $t+1,t+2,\dots$. The frameworks presented above, however, assume that unless nodes are in the new seed set $S_{t+1}$, they are unable to exert influence. Next, save~\cite{singh2021link}, each maintenance seeding method assumes that the future network topology is known, which is generally untrue in practice. In particular, sequential seeding strategies when the graph snapshots are unknown is an important and practically relevant open problem. Finally, Yang et al.~\cite{yang2017tracking} argue that identifying influential nodes is a separate task from influence maximization. For example, if a new user joins Twitter, they may want to follow the most influential users. Identifying these users is different from trying to maximize the spread of a product or idea on Twitter's network.

The node probing problem is a promising step toward practically relevant IM algorithms. Indeed, assuming that the network structure is unknown, except through probing, is much more realistic than the methods which assume complete topological information. These methods, however, treat the problem as a sequence of static IM tasks since the seed nodes are computed fresh at each time step. An interesting advance would be to leverage the previous seed set in computing the new seeds.

\section{Real world implementations}\label{sec:real}

A key focus of this review is understanding if existing methods are prepared to handle IM tasks ``in the field.'' To date, the literature on IM in real-world settings is scant. In this section, we highlight the existing studies and discuss some of the associated challenges. While these works assume that the network is static, the majority employ a sequential seeding strategy which is why we include it in our discussion of temporal IM. To our knowledge, there are no existing papers explicitly implementing IM algorithms on dynamic networks.

The most notable examples of applied IM comes from a series of papers by Yadav and Wilder~\cite{yadav2016using, yadav2017influence, yadav2018bridging, wilder2017uncharted, wilder2018end}. In these works, the goal is to maximize HIV awareness among homeless youth in large urban areas. This is a classic IM setting as homeless shelters can only train a small number of youth on HIV prevention, but hope that participants pass this information along to their friends to maximize awareness. The general problem setup is as follows. First, the social network of homeless youth is partially constructed. Then the homeless shelter chooses $k$ youth to participate in an intervention on HIV prevention. During the training, the youth reveal all of their one-hop friendships. The information is then given time to diffuse on the network (but this spread is unknown) before inviting $k$ more youth for training. This process continues for $T$ training rounds.

There are several key challenges to deploying IM algorithms in this setting. First, the complete social network of the homeless youth population is unknown, both in terms of nodes (youth) and links (friendships). Moreover, new information on the network structure is collected during the experiment as youth are trained and their friendship circle is elucidated. Second, youth may refuse and/or be unable to attend the training, meaning that seed nodes have a certain probability of remaining inactive. Lastly, quantifying the information spread on the network is highly non-trivial. Thus, this problem combines node probing, as the network structure is partially unknown before selecting a node to learn their social circle, and sequential seeding, where the nodes are activated over time. It differs from the standard node probing problem, however, in that nodes are chosen to optimize influence spread, rather than maximize topological information about the network; it differs from sequential seeding in that the influence spread is unknown when selecting the next seed nodes.

The first attempt to address these challenges comes from~\cite{yadav2016using}. Assuming the SI model\footnote{In the paper, Yadav et al.\ state that they use the IC model, but in terms of the notation of this paper, it falls under the SI classification} for diffusion, the authors construct the social network using Facebook friendships while inferring missing links using link prediction techniques~\cite{kim2011network}. They prove that the task of choosing $k$ seed nodes at each of the $T$ time steps is NP-hard and that it is impossible to achieve a $n^{-1+\epsilon}$ approximation of the optimal solution with an uncertain network. The problem is then recast as a Partially Observable Markov Decision Process (POMDP). By simulating the diffusion process, the nodes with the largest expected reward (influence spread) are selected for the seed set. In order for the method to handle real-world network sizes, the authors propose a divide-and-conquer approach. Their proposed method is one hundred times faster than existing methods while also yielding greater influence spread. In~\cite{yadav2016using}, the authors generalize the model by allowing for greater uncertainty in the influence and edge probabilities.

\begin{figure}
    \centering
    \includegraphics[scale=0.55]{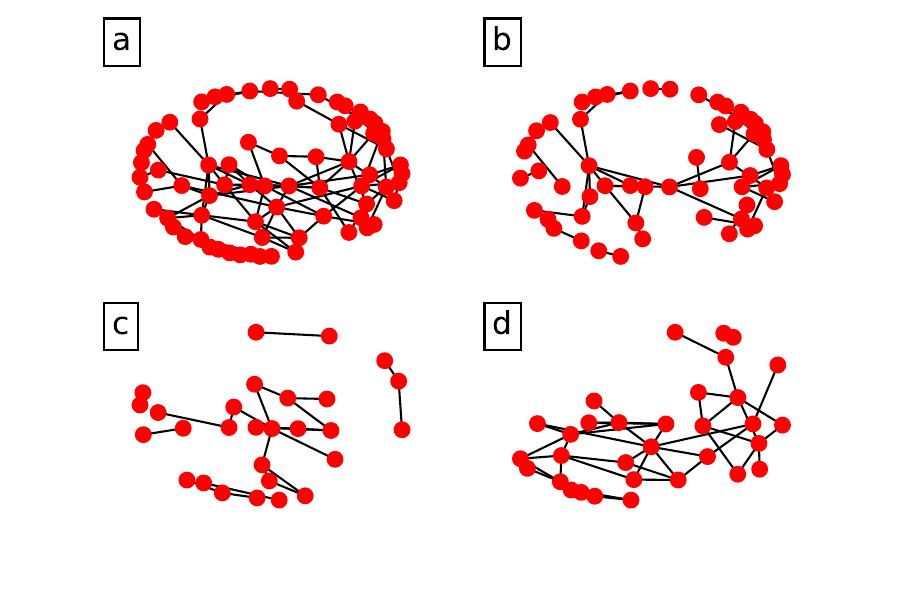}
    \caption{Homeless youth social network constructed using different methods. (a) All methods combined. (b) Self-reported edges. (c) Field observations. (d) Staff observations. Reproduced from \cite{wilder2018end} with permission of the author.}
    \label{fig:real}
\end{figure}

In~\cite{wilder2018end}, the authors focus on several practical considerations for this problem. First, the algorithm accounts for a non-zero probability that a seed set node remains inactive, i.e., the youth does not attend the training. The authors also address the network construction step by proposing a network sampling approach based on the {\it friendship paradox}~\cite{feld1991your}. This paradox says that, on average, a random node's neighbor has more friends than the original node. Thus, with a sampling budget of $M$ nodes, they first randomly sample $M/2$ nodes and then randomly sample one neighbor per node. This approach increases the likelihood that central (e.g., influential) nodes are sampled. Figure \ref{fig:real} shows the homeless youth social network constructed using different methods (reproduced with permission of the author). These four networks highlight the challenges of constructing the network for a hard-to-reach population as the topology varies greatly depending on the collection method (self-report, field observations and homeless shelter staff observations). Next, the authors assume that the influence propagation probability is unknown but modeled to maximize the worst-case ratio between the true spread and the estimated spread. Finally, the authors propose a greedy algorithm to select the optimal seed nodes and prove that it is guaranteed to output a solution within a factor of $(e-1)/(2e-1)$ of the optimal. In a real-world pilot study, by sampling only 15\% of the nodes, the proposed method achieved comparable spread compared to that if the entire network was known.

These methods are applied to the real-world task in~\cite{yadav2017influence, yadav2018bridging}. Some of the key questions considered are: Do the activated nodes actually pass their information along to others? Do the activated nodes give meaningful information about the social network? Can these algorithms do a better job of selecting seed nodes than an expert (social worker) can? To answer these questions, the authors implement the methods from~\cite{yadav2016using} and~\cite{wilder2017uncharted}. They also consider a baseline IM algorithm based on largest degrees. For each method, the authors recruit study participants, construct the network, activate nodes (via training) and conduct follow-ups to evaluate the final influence spread. The proposed methods in~\cite{yadav2016using} and~\cite{wilder2017uncharted} yielded much larger influence spreads than the degree-based method while also leading to a change in participants' behavior, i.e., increase in participants testing for HIV.
%This paper also highlighted the challenge of participants not attending training and the difficulties in comparing different methods when applied to different social networks. 

Lastly, we discuss an application of IM to an online setting. \cite{huang2022influence} consider a ``closed'' social network where posts are only shared with certain people rather than all of the users' connections. As a slight variation to the standard IM problem, the authors find the friends with which the user should share their information to maximize spread. In other words, the goal is to maximize the influence spread where users can only share the information with a limited subset of edges (neighbors). Indeed, this may be a more realistic diffusion mechanism for social networks as it is unlikely that someone would give equal effort to share information with each of their friends; rather, he/she would likely target a few specific people. The authors apply this to an online multiplayer game where each user is recommended friends to interact with, e.g., send gifts, game invitations, etc. The proposed method is compared with randomly selecting friends and yields a 5\% increase in click-through rate.

\section{Considerations and future directions}\label{sec:conc}

We concluded by sharing thoughts on the challenges associated with temporal IM as well as some of the important areas for future research.

\subsubsection*{Real-world implementations:} In Section \ref{sec:real}, we saw the litany of challenges facing a researcher trying to implement IM algorithms on real-world problems. We list a handful of questions that he/she must consider in applying these methods: What is the information diffusion mechanism? Can nodes be sequentially updated, or are they all activated at the start? Will seed nodes be activated with certainty? How long does the diffusion process continue? On what time scale is the network evolving? How long does it take to influence a node? Are the network dynamics changing rapidly? Does the future topology of the network need to be predicted? Is the network updated in an online setting or with standard snapshots? Is the true influence spread known? We look forward to many more IM implementations in real-world applications.

\subsubsection*{Single seeding methods:} In Section \ref{sec:one}, we discussed several methods for the single seeding temporal IM problem. There were only five papers, however, and Aggarawal et al.~\cite{aggarwal2012influential}, Osawa and Murata~\cite{osawa2015selecting} and Erkol et al.~\cite{erkol2020influence} all proposed similar solutions. Thus, there is still much room for research on this problem. Recently, graph neural networks (GNN) were applied to the static IM problem~\cite{qiu2018deepinf, tian2020deep, kumar2022influence} and may also find success in the temporal setting. 

\subsubsection*{Ex ante vs.\ ex post:} Most methods proposed in Section \ref{sec:one} assume that the entire topology of the dynamic network is known, even though this is unrealistic in many situations.~\cite{yanchenko2023link} yielded promising results for {\it ex ante} IM, but more work is certainly needed. 

\subsubsection*{Impact of time:} In dynamic networks with diffusion, there is a highly intricate relationship between the structural evolution and influence diffusion. This must be carefully accounted for in the IM problem similar to Gayraud et al.~\cite{gayraud2015diffusion} and~\cite{albano2013matter}. The impact of time scales, aggregation, diffusion times, and diffusion mechanisms deserves further study.

\subsubsection*{Online setting:} Related to the previous point is IM in the online setting, where nodes and edges come and go continuously. In real-world applications, it may not be obvious how or when to aggregate the network so it becomes more natural to consider online updates. Most methods, however, require that the network is aggregated into graph snapshots. This aggregation inherently loses information, such as when the link appeared/disappeared and the persistence of the edge. More methods like Ohsaka et al.~\cite{ohsaka2016dynamic} can be developed to address this challenge.

\subsubsection*{Model mis-specification:} A pertinent challenge for applied IM is selecting the diffusion model. For diseases, the SIR model is sensible since infected persons can infect other nodes for as long as they are infected. On the other hand, for HIV awareness among homeless youth, it is unlikely that someone would attempt to influence all of his/her friends indefinitely. Thus, choosing an appropriate diffusion model is crucial. But what are the effects on influence spread if the model is misspecified? In~\cite{aral2018social}, the authors study this for static IM and find that standard diffusion models grossly underestimate the influence spread of more realistic models. This is likely only compounded in temporal networks where the topology also varies.
    
\subsubsection*{Uncertainty estimates of seed nodes:} The majority of temporal IM algorithms output the optimal seed nodes to achieve maximal influence spread. But are there other seed sets that would yield a comparable spread? In other words, is the objective function ``flat'' in the sense that many seed sets yield comparable spread? An interesting avenue of research would be deriving a measure of uncertainty for optimal seed sets.

\subsubsection*{Influence minimization:} A related problem to IM is that of influence {\it minimization} in which seed nodes are ``vaccinated'' to stop the spread of influence on the network. This problem arises in rumor diffusion and epidemiological settings~\cite{minwang2013negative, minwang2020efficient, minyang2019influence, minwang2017drimux} and may lead to interesting philosophical questions. For example, in the vaccine campaign against COVID-19, vaccines were first administered to the most vulnerable populations, e.g., elderly. Thus, seed nodes were chosen based on vulnerability. In an influence minimization schema, however, the most active and/or social people would likely receive the vaccine first to minimize the spread between groups. These opposing goals lead to challenging decisions both ethically and politically.

\ifCLASSOPTIONcompsoc
  \section*{Acknowledgments}
\else
  \section*{Acknowledgment}
\fi

This work was conducted while EY was on a JSPS Predoctoral Fellowship for Research in Japan (Short-term Program). PH was supported by JSPS KAKENHI Grant Number JP 21H04595.

\ifCLASSOPTIONcaptionsoff
  \newpage
\fi



%\bibliographystyle{IEEEtran}

\bibliographystyle{IEEEtran}
\bibliography{refs}


\begin{IEEEbiography}{Eric Yanchenko}
received a BS in Mathematics and Physics from The Ohio State University, Columbus, Ohio, USA. He is currently a PhD candidate in Statistics at North Carolina State University, Raleigh, North Carolina, USA and a Japan Society for Promotion of Science (JSPS) Short-term Fellow at Tokyo Institute of Technology, Tokyo, Japan. His research interests include hypothesis testing for meso-scale structures on graphs, eliciting prior distributions for scale parameters in hierarchical Bayesian models, and influence maximization. 
\end{IEEEbiography}

\begin{IEEEbiography}{Tsuyoshi Murata} received his bachelor's degree from the Department of Information Science at The University of Tokyo, Tokyo, Japan and his PhD from Tokyo Institute of Technology, Tokyo, Japan, where he is currently a full professor in the School of Computing. His research interests are in artificial intelligence, network science, machine learning and social network analysis.

\end{IEEEbiography}

\begin{IEEEbiography}{Petter Holme} received a PhD in Theoretical Physics from Umea University, Umea, Sweden. He has served as a professor at Sungkyunkwan University, Seoul, Korea and the Institute of Innovative Research, Tokyo Institute of Technology, Tokyo, Japan. He is currently a full professor of Computer Science at Aalto University, Espoo, Finland and maintains an affiliation with the Center for Computational Social Science, Kobe University, Kobe, Japan. His research focuses on large-scale structures in society, technology, and biology.
\end{IEEEbiography}


%\vfill


\end{document}